\documentclass[prl,aps,showpacs,floatfix,twocolumn]{revtex4}

\usepackage{amsmath,amssymb,amsxtra,bm,bbm}
\usepackage[dvips]{graphicx}

\begin{document}

\title{Theory of Scanning Tunneling Spectroscopy of Magnetic Adatoms in
Graphene}

\author{Bruno Uchoa$^{1}$, Ling Yang$^{2}$, S.-W. Tsai$^{2}$, N.~M.~R.
Peres$^{3}$, and A.~H. Castro Neto$^{4}$}

\affiliation{$^{1}$Department of Physics$\mbox{,}$ University of Illinois at
Urbana-Champaign, 1110 W. Green St, Urbana, IL, 61801, USA}

\affiliation{$^{2}$Department of Physics and Astronomy, University of California,
Riverside, CA, 92521,USA}

\affiliation{$^{3}$Centro de F\'{\i}sica e Departamento de F\'{\i}sica, Universidade
do Minho, P-4710-057, Braga, Portugal}

\affiliation{$^{4}$Department of Physics, Boston University, 590 Commonwealth
Avenue, Boston, MA 02215, USA}

\date{\today}

\begin{abstract}
We examine theoretically the signatures of magnetic adatoms in graphene
probed by scanning tunneling spectroscopy (STS). When the adatom hybridizes
equally with the two graphene sublattices, the broadening of the local
adatom level is anomalous and can scale with the \emph{cube} of the
energy. In contrast to ordinary metal surfaces, the adatom local moment
can be suppressed by the proximity of the probing scanning tip. We
propose that the dependence of the tunneling conductance on the distance
between the tip and the adatom can provide a clear signature for the
presence of local magnetic moments. We also show that tunneling conductance
can distinguish whether the adatom is located on top of a carbon atom
or in the center of a honeycomb hexagon. 
\end{abstract}


\maketitle
Graphene is a two dimensional sheet of carbons whose remarkable electronic
properties derive from the presence of electronic excitations that
behave as chiral Dirac quasiparticles\cite{novo3}. Although clean
bulk graphene may not be magnetic, there is a rich variety of possibilities
for magnetism when adatoms are added on top of graphene. As an open
surface, the use of scanning tunneling microscopy (STM) probes \cite{stolyarova}
opens the possibility of controlling the position of adatoms with
atomic precision\cite{eigler} and at the same time switching the
magnetic local moments on and off by gating\cite{uchoa,sengupta}. 

One of the challenges for the manipulation of local moments is about
detection: how can one reliably identify a local magnetic moment at
room temperature. Unlike ordinary metal surfaces, due to the low density
of states (DOS), an adatom localized level can hybridize strongly
with the STM tip. We propose that the dependence of the adatom STM
differential conductance (DC) with the distance to a \emph{non-magnetic
}STM tip, in the limit of small separation, can provide an experimental
signature for the presence of local moments \emph{above} the Kondo
temperature\cite{fradkin90}. Furthermore, because the electrons
in graphene carry different sublattice quantum numbers, we show that
when the adatom sits in the center of the honeycomb hexagon {[}see
Fig.1(b)], destructive interference between the different tunneling
paths changes substantially the form of the Fano factor\cite{fano}
and the shape of the DC curves compared to the case where the adatom
is located on the top of a carbon atom {[}Fig.1(a)]. This effect allows
the use of STM to characterize adatoms and defects in graphene, including
substitutional impurities in single and double vacancies\cite{kras}.

\begin{figure}[b]
\begin{centering}
\vspace{-0.2cm}
 \includegraphics[scale=0.27]{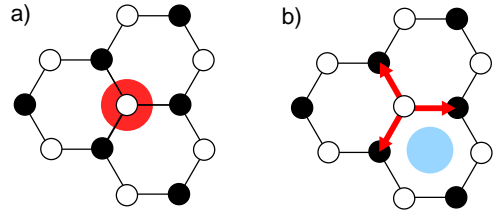}\vspace{-0.1cm}

\par\end{centering}

\caption{{\small Two adatom positions in graphene: a) asymmetric case, on top
of a carbon atom, when the adatom (large circle) hybridizes with one
sublattice and b) symmetric case, when the adatom is located at the
center of a hexagon. Red arrows: nearest neighbor vectors.}}

\end{figure}

Our starting point is the free Hamiltonian of the graphene/adatom/tip
system: ${\cal H}={\cal H}_{g}$$+{\cal H}_{f}$$+{\cal H}_{c}$.
${\cal H}_{g}$ is the tight-binding Hamiltonian for graphene: ${\cal H}_{g}=-t\sum_{\langle ij\rangle}\sum_{\sigma}a_{\sigma}^{\dagger}(\mathbf{R}_{i})b_{\sigma}(\mathbf{R}_{j})+{\rm h.c.},$
where $a,b$ are the fermionic operators for sublattices $A$ and
$B$, respectively ($t\sim2.8$ eV), and $\sigma=\uparrow\downarrow$
is the spin. In momentum space, \begin{equation}
{\cal H}_{g}=-t\sum_{\sigma\mathbf{k}}\left[\phi(\mathbf{k})a_{\mathbf{k}\sigma}^{\dagger}b_{\mathbf{k}\sigma}+\phi^{*}(\mathbf{k})b_{\mathbf{k}\sigma}^{\dagger}a_{\mathbf{k}\sigma}\right],\label{eq:Hg}\end{equation}
where $\phi(\mathbf{k})=\sum_{i=1}^{3}\mbox{e}^{i\mathbf{k}\cdot\mathbf{a}_{i}}$,
$\mathbf{a}_{1}=\hat{x}$, $\mathbf{a}_{2}=-\hat{x}/2+\sqrt{3}\hat{y}/2$,
and $\mathbf{a}_{3}=-\hat{x}/2-\sqrt{3}\hat{y}/2$ are the lattice
nearest neighbor vectors. ${\cal H}_{c}=\sum_{\mathbf{k}}\epsilon_{\mathbf{k}}c_{\mathbf{k},\sigma}^{\dagger}c_{\mathbf{k}\sigma}$
is the effective Hamiltonian for the $c$ tip electrons, with $\epsilon_{\mathbf{k}}=(k^{2})/2m^{*}-\epsilon_{D}$,
where $m^{*}$ is the effective electronic mass, and $\epsilon_{D}$
is the energy at the bottom of the tip band with respect to the Dirac
point, and ${\cal H}_{f}=\sum_{\sigma}\epsilon_{0}\, f_{\sigma}^{\dagger}f_{\sigma}$
is the Hamiltonian of the $f$ electrons at the local level with energy
$\epsilon_{0}$. The Coulomb energy, $U$, for double occupancy of
the local level is described by a Hubbard term: ${\cal H}_{U}=Uf_{\uparrow}^{\dagger}f_{\uparrow}f_{\downarrow}^{\dagger}f_{\downarrow}$.
Since we are only interested in the magnetic state of the adatom (we
do not include the Kondo effect and hence our theory is valid above
the adatom Kondo temperature, $T_{K}$), in what follows we use Anderson's
mean field decomposition\cite{uchoa}: ${\cal H}_{U,{\rm MF}}=U\sum_{\sigma}n_{\sigma}f_{-\sigma}^{\dagger}f_{-\sigma}-Un_{\downarrow}n_{\uparrow},$
where $n_{\sigma}=\langle f_{\sigma}^{\dagger}f_{\sigma}\rangle$
is the average occupation of the level. Hence, at the mean field level
we write: ${\cal H}_{f,{\rm MF}}=\sum_{\sigma}\epsilon_{\sigma}\, f_{\sigma}^{\dagger}f_{\sigma}\,,$
where $\epsilon_{\sigma}=\epsilon_{0}+Un_{-\sigma}$ is the renormalized
level energy.

In graphene the adatoms can be localized at different positions in
the honeycomb lattice\cite{kras}. Here we consider two cases where
the adatom is either placed on top of a carbon atom, in which case
the sublattice symmetry is locally broken, or the adatom is located
in the center of an hexagon without symmetry breaking. In the first
case, assuming the adatom to be on the $B$ sublattice, we have: ${\cal H}_{V,{\rm AS}}\!=\! V\sum_{\sigma}f_{\sigma}^{\dagger}b_{\sigma}(0)+{\rm h.c.}$.
In the second case we have: ${\cal H}_{V,{\rm S}}\!=\!\sum_{\sigma}\sum_{i=1}^{3}V\left[a_{\sigma}^{\dagger}(\mathbf{a}_{i})+b_{\sigma}^{\dagger}(-\mathbf{a}_{i})\right]f_{\sigma}+h.c.$.
In momentum space, these two terms can be written as: \begin{equation}
{\cal H}_{V,{\rm AS}}=V\sum_{\mathbf{p}\sigma}b_{\mathbf{p}\sigma}^{\dagger}f_{\sigma}+{\rm h.c.}\label{eq:bf}\end{equation}
 in the asymmetric case, and \begin{equation}
{\cal H}_{V,{\rm S}}=V\sum_{\mathbf{p}\sigma}\left[\phi(\mathbf{p})b_{\mathbf{p}\sigma}^{\dagger}+\phi^{*}(\mathbf{p})a_{\mathbf{p}\sigma}^{\dagger}\right]f_{\sigma}+{\rm h.c.}\label{eq:baf}\end{equation}
 in the symmetric one \cite{note2}.

In the presence of an STM tip, there are two additional hopping terms:
${\cal H}_{f-c}=V_{c}\,\sum_{\sigma}f_{\sigma}^{\dagger}c_{\sigma}(\mathbf{r})+{\rm h.c.}$,
where $V_{c}$ is the effective tunneling energy between the tip and
the adatom, and ${\cal H}_{g-c}=\sum_{\sigma,i}t_{c,i}\left[a_{\sigma}^{\dagger}(\mathbf{R}_{i})+b_{\sigma}^{\dagger}(\mathbf{R}_{i})\right]c_{\sigma}(\mathbf{R}_{i}-\mathbf{r})+{\rm h.c.}$
where $t_{c,i}$ is the hopping energy between the tip and each carbon,
$\mathbf{r}=(R,z)$ is the position of the tip with respect to the
adatom, $R$ is the in-plane distance from the adatom to the center
of the tip, and $z$ is the out-of-plane distance between them. The
fact that the adatom sits a few angstroms above the graphene plane
is accounted by the exponential $z$ dependence of the tunneling energies.
In momentum space we have: ${\cal H}_{f-c}=V_{c}(r)\sum_{\sigma\mathbf{p}}f_{\sigma}^{\dagger}c_{\sigma\mathbf{p}}+{\rm h.c.},$
and ${\cal H}_{g-c}=\sum_{\sigma\mathbf{k}\mathbf{p}}t_{c,\mathbf{p}}(\mathbf{r})\left[\, a_{\sigma\mathbf{p}}^{\dagger}+b_{\sigma\mathbf{p}}^{\dagger}\right]c_{\sigma\mathbf{k}}+{\rm h.c.},$
where $t_{c,\mathbf{p}}(\mathbf{r})=t_{c}(z)\mbox{e}^{-i\mathbf{p}\cdot\mathbf{R}}$
after averaging over the position of the carbon sites below the impurity.

In the absence of the tip, after diagonalization of the Hamiltonian
${\cal H}_{g}+{\cal H}_{f}+{\cal H}_{V}$ in Eq. (\ref{eq:Hg})$-$(\ref{eq:baf}),
the retarded Green's function of the $f$ electrons $G_{ff,\sigma}(\tau)=-\langle T[f(\tau)f^{\dagger}(0))]\rangle$
is given by: \begin{equation}
G_{ff,\sigma}^{R}(\omega)=\left[\omega-\epsilon_{\sigma}-\Sigma_{ff}(\omega)\right]^{-1}\label{eq:Gff}\end{equation}
 where $\Sigma_{ff}(\omega)$ is the self-energy of the localized
electrons due to the hybridization with the electrons in graphene.
As in the usual Anderson impurity problem\cite{anderson61}, the
formation of local moments is defined by the occupation of the localized
level $n_{\sigma}$ for up and down spins, $n_{\sigma}=-(1/\pi)\int_{-D}^{\mu}\mbox{d}\omega\,\mbox{Im}G_{ff,\sigma}^{R}(\omega),$
where $\mu$ is the chemical potential ($D\sim7$ eV is an energy
cut-off).

In the asymmetrical case {[}see Eq.(\ref{eq:bf})], $\Sigma_{ff}^{{\rm AS}}(\omega)=V^{2}\sum_{\mathbf{p}}G_{bb,\mathbf{p}}^{0\, R}(\omega)$,
where $G_{bb}^{0}(\tau)=-\langle T[b(\tau)b^{\dagger}(0)]\rangle$
is the diagonal component of the bare graphene Green's function, $G_{xy}^{0}$,
with $x,y=a,b$: \begin{equation}
G_{xy,\mathbf{k}}^{0\, R}(\omega)=\frac{\omega\sigma_{xy}^{0}-t\sigma_{xy}^{1}\mbox{Re}\phi(\mathbf{k})+t\sigma_{xy}^{2}\mbox{Im}\phi(\mathbf{k})}{\omega^{2}-t^{2}|\phi(\mathbf{k})|^{2}+i0^{+}\mbox{sign}(\omega)}\,,\label{eq:G}\end{equation}
 where $\sigma^{j}$ ($j=1,2)$ are off-diagonal Pauli matrices ($\sigma_{ba}^{2}=\sigma_{ab}^{2*}=i$),
and $\sigma^{0}$ is the identity matrix. We calculate the self-energy
within the linearized theory, where the energy around the $K$ ($K^{\prime}$
) point is given by: $\epsilon_{\pm}(k)=\pm v_{F}k$, where $v_{F}\approx10^{6}$
m/s is the Fermi velocity. It reads\cite{uchoa}: \begin{equation}
\Sigma_{ff}^{{\rm AS}}(\omega)=-(\omega\Delta/\pi)\ln\left(|D^{2}-\omega^{2}|/\omega^{2}\right)-i\Delta|\omega|\theta(D-|\omega|)\,,\label{eq:sigma_A}\end{equation}
 where $\Delta=\pi(V/D)^{2}$ is the dimensionless hybridization parameter.
The imaginary part of (\ref{eq:sigma_A}) gives the broadening of
the level due to the hybridization of the adatom, and it is proportional
to the DOS of the host.

In the symmetric case, however, the hybridization of the $f$ electrons
is mediated by the virtual hopping of the electrons in graphene into
a {}``ghost'' site located below the adatom, in the center of the
hexagon. In this process, as the electrons hop in and out from the
adatom, they gain an additional phase that leads to interference between
the different quantum mechanical paths involving the six adatom neighboring
sites on both sublattices. The self-energy in this case involves also
off-diagonal terms of (\ref{eq:G}): \begin{equation}
\Sigma_{ff}^{{\rm S}}(\omega)=V^{2}\sum_{\mathbf{p}}\left[\phi(\mathbf{p})\Theta_{a,\mathbf{p}}(\omega)+\phi^{*}(\mathbf{p})\Theta_{b,\mathbf{p}}(\omega)\right],\label{eq:Sigma_S}\end{equation}
 where $\Theta_{x,\mathbf{p}}(\omega)=\phi(\mathbf{p})G_{xb,\mathbf{p}}^{0\, R}(\omega)+\phi^{*}(\mathbf{p})G_{xa,\mathbf{p}}^{0\, R}(\omega)$.
In the linearized theory, Eq. (\ref{eq:Sigma_S}) gives\cite{note}:
\begin{equation}
\Sigma_{ff}^{{\rm S}}(\omega)=-\omega\left[Z^{-1}(\omega)-1\right]-2i\Delta(|\omega|^{3}/t^{2})\theta(D-|\omega|)\,,\label{eq:sigma_B}\end{equation}
 where $Z^{-1}=1-(1/\omega)\mbox{Re}\Sigma_{ff}^{{\rm S}}(\omega)=1+2\Delta/(\pi t^{2})[D^{2}+\omega^{2}\ln\left(|D^{2}-\omega^{2}|/\omega^{2}\right)]$
gives the quasiparticle residue. The imaginary part of $\Sigma_{ff}^{{\rm S}}$
gives rise to an anomalous broadening of the adatom level that scales
with $|\omega|^{3}/t^{2}$, suppressing strongly the hybridization
when $|\epsilon_{0}|\ll t$.

In the perturbative regime where $V_{c},t_{c}$ are small compared
to $V$, the inclusion of the tip leads to an additional renormalization
of the $f$ electrons Green's function, $\Sigma_{ff}(\mathbf{r},\omega)=\Sigma_{ff}(\omega)+\Sigma_{ff}^{(1)}(\mathbf{r},\omega)$,
where \begin{equation}
\Sigma_{ff}^{(1)}(\mathbf{r},\omega)=V_{c}(\mathbf{r},\omega)\bar{V}_{c}(-\mathbf{r},\omega)\sum_{\mathbf{k}}G_{cc,\mathbf{k}}^{0\, R}(\omega)\,.\label{eq:Sigma_(1)}\end{equation}
 $G_{cc,\mathbf{k}}^{0\, R}(\omega)=\left[\omega-\epsilon(\mathbf{k})+i0^{+}\right]^{-1}$
is the retarded Green's function of the $c$-electrons, $G_{cc,\mathbf{k}}^{0}(\tau)=-\langle T[c_{\mathbf{k}}(\tau)c_{\mathbf{k}}^{\dagger}(0)]\rangle$,
whereas $V_{c}(\mathbf{r},\omega)$ is the renormalized tunneling
energy between the tip and the adatom, namely \begin{equation}
V_{c}^{AS}\negmedspace(\mathbf{r},\omega)=V_{c}\!+\! Vt_{c}(z)\!\left[G_{ab}^{0\, R}(\mathbf{R},\omega)\!+\! G_{bb}^{0\, R}(\mathbf{R},\omega)\right]\label{eq:V_A}\end{equation}
 for the asymmetric case, where $\bar{V}_{c}(\mathbf{r},\omega)$
follows from the exchange $G_{xb}^{0}\to G_{bx}^{0}$, and \begin{equation}
V_{c}^{S}(\mathbf{r},\omega)=V_{c}+Vt_{c}(z)\left[\Theta_{a}(\mathbf{R},\omega)+\Theta_{b}(\mathbf{R},\omega)\right]\label{eq:V_S}\end{equation}
 in the symmetric one ($V^{S}\equiv\bar{V}^{S}$). In our notation,
$\Theta(\mathbf{R})=\sum_{\mathbf{p}}\mbox{e}^{i\mathbf{p}\cdot\mathbf{R}}\Theta_{x,\mathbf{p}}$
is the Fourier transform of $\Theta_{x}$. $\Sigma_{ff}^{(1)}$ in
Eq. (\ref{eq:Sigma_(1)}) can be easily computed assuming an effective
band width, $\alpha_{D}$, for the $c$ electrons in the tip.

\begin{figure}
\begin{centering}
\includegraphics[scale=0.39]{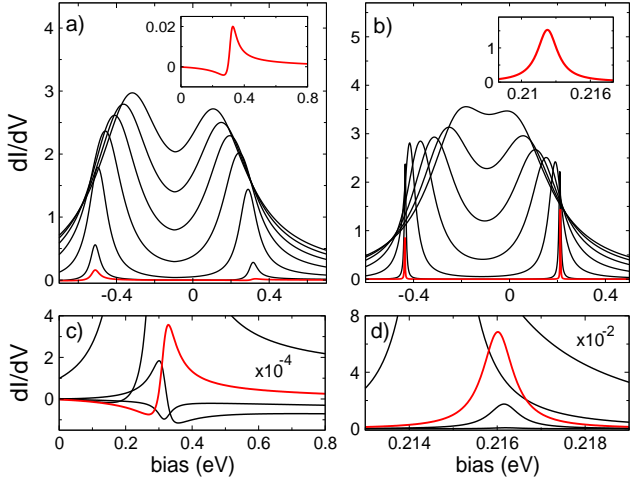} 
\par\end{centering}

\caption{{\small Adatom induced DC, $\mathcal{G\equiv}\mbox{d}I/dV$, versus
bias when the adatom sits (left) on top of carbon and (right) in the
center of the hexagon. See details in the text.  a), b) $t_{c}=0.15$eV
and $V_{c}/t_{c}=1.6$, $1.5$, $1.35$, $1.1$, $0.7$, $0.25$ and
$0.1$ (inset), from top to bottom. c), d) $t_{c}=0.02$eV and $V_{c}/t_{c}=$
$0.7$, $0.25$, $0.1$, $0.05$ and $0.01$. $V_{c}/t_{c}=0.1$ for
all curves in red.{}}{\footnotesize{} }}

\end{figure}

The adatom induced DC is $\mathcal{G}(\mathbf{r},\omega)\propto\rho_{c}(\omega)\rho_{f-c}(\mathbf{r},\omega)$,
where $\rho_{c}$ is the DOS of the tip, and $\rho_{f-c}(\mathbf{r},\omega)=-(1/\pi)\mbox{Im}\left[V_{c}(\mathbf{r},\omega)\sum_{\sigma}G_{ff,\sigma}(\omega)\bar{V}_{c}(-\mathbf{r},\omega)\right]$
is the analogue of the $f$ electron DOS, which contains renormalized
tunneling matrix elements between the tip and the adatom\cite{plihal}.
In a more standard form, the DC is \begin{equation}
\mathcal{G}(\mathbf{r},\omega)=2\pi e\rho_{c}\, t_{c}^{2}\,\tilde{\rho}\sum_{\sigma}\frac{q\bar{q}-\gamma\bar{\gamma}+(q\bar{\gamma}+\bar{q}\gamma)\xi_{\sigma}}{\xi_{\sigma}^{2}+1}\,,\label{eq:G1}\end{equation}
 where $e$ is the electron charge, $\omega=\mu+eV$, with $eV$ the
applied bias, $\tilde{\rho}(\mathbf{r},\omega)=-\rho^{2}(\omega)/\mbox{Im}\Sigma_{ff}(\mathbf{r},\omega)$,
where $\rho(\omega)$ is the graphene DOS per spin, and $\xi_{\sigma}(\mathbf{r},\omega)=-[\omega-\epsilon_{\sigma}-\mbox{Re}\Sigma_{ff}(\mathbf{r},\omega)]/\mbox{Im}\Sigma_{ff}(\mathbf{r},\omega),$
$q(\mathbf{r},\omega)=\mbox{Re}V_{c}(\mathbf{r},\omega)/[\pi t_{c}(z)V\rho(\omega)]$
and $\gamma(\mathbf{r},\omega)=-\mbox{Im}V_{c}(\mathbf{r},\omega)/[\pi t_{c}(z)V\rho(\omega)]$
are the Fano parameters\cite{fano}. In contrast with usual metal
surfaces, the Fano factor $q$ has a live dependence with the bias
in graphene. The conjugate forms $\bar{q}(-\mathbf{r},\omega)$ and
$\bar{\gamma}(-\mathbf{r},\omega)$ are defined by the conjugate tunneling
matrix element $\bar{V}_{c}(-\mathbf{r},\omega)$. In the particular
case when the tip is above the adatom (\textbf{$R=0$}), $q=\bar{q}$
and $\gamma=\bar{\gamma}$ and $q^{AS}=[V_{c}+(t_{c}(z)/V)\mbox{Re}\Sigma_{ff}^{AS}(\omega)]/\left[\pi t_{c}(z)V\rho(\omega)\right],$
where $\Sigma_{ff}^{AS}(\omega)$ is defined by Eq.(\ref{eq:sigma_A}),
and $\gamma^{AS}=1$. In the case where the adatom and the tip are
on top of each other in the center of the hexagon, destructive interference
leads to cancellation of the perturbative corrections on the tunneling
matrix element in Eq.(\ref{eq:V_S}), and the Fano parameter simplifies
to $q^{S}=V_{c}/\left[\pi t_{c}(z)V\rho\right]$ and $\gamma^{S}=0$.
For adatoms with $d$ and $f$-wave orbitals, the cancellation is
not exact. %
\begin{figure}[b]
\begin{centering}
\includegraphics[scale=0.45]{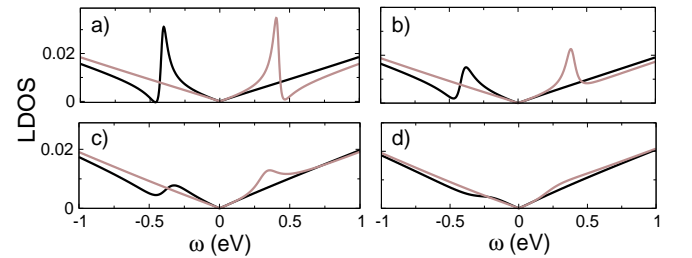} 
\par\end{centering}

\caption{{\small Graphene LDOS at the adatom site (top carbon case) for $V_{c}/t_{c}=0.1,\,0.7$,
$1.1$ and $1.6$, from (a) to (d) ($t_{c}=0.15$eV). Black curve:
$n_{\uparrow}$; brown: $n_{\downarrow}$. Total LDOS: $n_{\uparrow}+n_{\downarrow}$}{\footnotesize .}}

\end{figure}

The shape of the Fano resonances in the DC curves is driven by the
ratio $q/\gamma$. When $q/\gamma\gg1$, the DC curve shows a pronounced
peak, whereas in the opposite regime, $q/\gamma\ll1$ one expects
a dip. For a set of parameters $V=1$eV, $U=1$eV, $\alpha_{D}=4$
eV, $\epsilon_{D}=2$eV, $\mu=0.1$eV and $\epsilon_{0}=-0.5$eV,
in the asymmetric case, for $V_{c}/t_{c}=0.1$, the red curve shown
in the inset of Fig. 2(a) has a small dip, which is suppressed when
$V_{c}/t_{c}\gtrsim0.2$. In contrast, the curves shown in Fig.2(b)
for the symmetric case have a well pronounced peak for all finite
values of $q^{S}$ {[}see inset of Fig.2(b)], reflecting the fact
that $q^{S}/\gamma^{S}$ is always large ($\gamma^{S}=0$). Fig. 2(a),(b)
and 2(c),(d) compare the features of the positive bias resonance for
$t_{c}=0.15$eV and $t_{c}=0.02$ eV, respectively. All red curves
in Fig.2 correspond to $V_{c}/t_{c}=0.1$. Increasing this ratio from
$V_{c}/t_{c}=0.25$ up to $1.6$, the DC curves show two strongly
pronounced peaks indicating the position of the two magnetic Fano
resonances at $\epsilon_{0}+n_{\uparrow}U$ and $\epsilon_{0}+n_{\downarrow}U$.
For $V_{c}/t_{c}<0.1$, in the asymmetric case, Fig.2(c) shows an
inversion in the structure of the resonance ($V_{c}/t_{c}=0.01$)
for positive bias. The shape of the DC curves for small $q/\gamma$
{[}see insets of Fig.2] agrees with a recent STM measurement of the
Kondo peak in graphene\cite{manoharan}.

\begin{figure}
\begin{centering}
\includegraphics[scale=0.3]{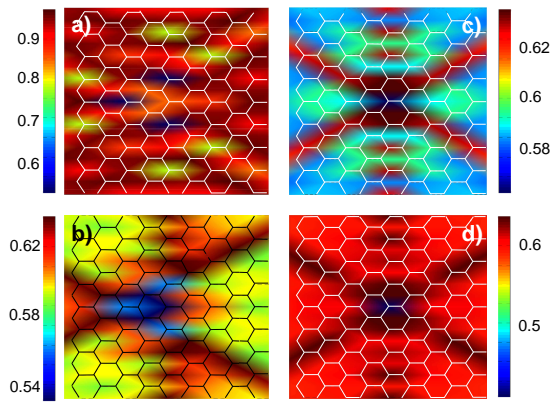} 
\par\end{centering}

\caption{{\small Integrated LDOS maps around the adatom (center) in the asymmetric
case, with the adatom on top of the carbon. On the left (right): scans
for the opposite (same) sublattice of the adatom: (a),(c) without
($t_{c}=V_{c}=0)$ and (b),(d) with the tip ($t_{c}=0.2$eV, $V_{c}=0.02$eV).}}

\end{figure}

The decrease in the separation of the peaks with increasing $V_{c}$
reflects the suppression of the local moment by the proximity of the
STM tip. In particular, in the symmetric case {[}Fig. 2(b)], the hybridization
of the adatom with graphene is weaker than in the asymmetric one,
making local moment much more sensitive to the STM tip. The difference
is indicated clearly by the separation of the peaks for large $V_{c}$
and also by the width of the peaks as $V_{c}$ goes to zero. In the
asymmetric case (Fig. 2(a)), the peaks remain broad at small $V_{c}$
whereas in the other case their width collapses much faster, reflecting
their anomalous broadening $\propto|\omega|^{3}/t^{2}$ {[}see Eq.
(\ref{eq:sigma_B})]. In the opposite limit, for $V_{c}$ large enough,
the two DC magnetic peaks eventually merge on top of each other, destroying
the local moment completely. The merging of the peaks happens much
earlier in the symmetric case {[}Fig. 2(b)] than in the asymmetric
one. We note that for large $V_{c}$, fluctuations drive the DC away
from equilibrium (although still in the perturbative regime for $V_{c}\ll V$),
invalidating the strict applicability of Eq. (\ref{eq:G1}). The main
effect of fluctuations, however, is to further suppress the local
moment, and in this sense the equilibrium calculation may be regarded
as a conservative estimate for the main effect, the suppression of
the local moment by the STM tip. As the tip separation to the adatom
becomes progressively small the DC peaks can shift strongly: in that
case, the peak on the right will \emph{red shift} and eventually cross
the experimentally accessible bias window around the Fermi level,
providing an experimental signature for the presence of the local
moment with a \emph{non-magnetic} tip, regardless the presence of
the Kondo peak. A significant suppression of the local moment by the
metallic tip is harder for adatoms with a very large $U$, such as
cobalt, which show a large local moment when hybridized with metals,
but may be easily achieved in adatoms which are not usually magnetic
and exhibit a local moment in graphene\cite{uchoa}.

The Fano resonance also generates magnetic peaks in the LDOS around
the adatom. The plots in Fig.3 show the evolution of those peaks with
the increase of $V_{c}$ in the asymmetric case, for up and down spins.
As the level is broadened by the tip, the height of the peaks clearly
collapses {[}Fig. 3(d)]. 

In the case where the adatom is on top of the carbon, the sublattice
asymmetry in the LDOS provides another STM signature that identifies
the position of the adatom in the lattice. The LDOS for the asymmetric
case can be computed perturbatively for small $t_{c}$ and $V_{c}$,
\begin{eqnarray}
\rho_{x,\sigma}^{AS}(\mathbf{r},\omega) & = & -1/(\pi V^{2})\left.\mbox{Im}\Sigma_{ff,\sigma}^{AS}(z,\omega)\right|_{V_{c}=0}-(1/\pi)\times\nonumber \\
 &  & \mbox{Im}\left[\Gamma_{x}(\mathbf{r},\omega)G_{ff,\sigma}^{R}(\omega)\bar{\Gamma}_{x}(-\mathbf{r},\omega)\right],\mbox{\qquad}\label{eq:rho_A}\end{eqnarray}
 where $x=b$ for the same sublattice of the adatom and $x=a$ for
the opposite one. The first term is the renormalized LDOS in the absence
of the adatom and \[
\Gamma_{x}(\mathbf{r},\omega)\!=\! VG_{xb}^{0R}(\mathbf{R},\omega)+\frac{t_{c}(z)G_{aa}^{0\! R}(0,\omega)}{\bar{V}_{c}^{\! AS}(-\mathbf{r},\omega)}\Sigma_{ff}^{(1)AS}(\mathbf{r},\omega)\]
 contains the interference effects due to the interplay of the adatom
and the tip in graphene. $\bar{\Gamma}_{x}$ in Eq.(\ref{eq:rho_A})
follows by exchanging $G_{xb}^{0}$ by $G_{bx}^{0}$ and $V_{c}^{AS}$
by $\bar{V}_{c}^{AS}$ {[}see Eq.(\ref{eq:V_A})]. The STM topography
maps computed from integration of Eq. (\ref{eq:rho_A}) in energy
(see Fig.4) clearly show the asymmetry between the two sublattices
(the adatom has three nearest neighbors and six next nearest neighbors).
Fig. 4(a),(b) display the integrated LDOS for the opposite sublattice
of the adatom, which has a lower point group symmetry, while Fig.4(c),(d)
display similar maps for the same sublattice of the adatom, with and
without interference effects from the tip. 

In conclusion, we have derived the fingerprints for Fano resonances
of magnetic adatoms in graphene. We have shown the signatures in the
DC curves that identify the position of the adatom and possibly the
presence of local moments, away from the Kondo regime.

We acknowledge E. Fradkin, K. Sengupta, H. Manoharan and E. Andrei
for discussions. BU acknowledges partial support from U.S. Department
of Energy under the grant DE-FG02-91ER45439 at University of Illinois,
and the Aspen Center of Physics, where this work started. SWT acknowledges
support from UC-Lab FRP under award number 09LR05118602. AHCN acknowledges
the partial support of the U.S. Department of Energy under the grant
DE-FG02-08ER46512. During the preparation of this paper we became
aware of a related work \cite{sengupta2}.


\begin{thebibliography}{10}
\bibitem{novo3}K. S. Novoselov et. al., Nature \textbf{438}, 197
(2005); Y. Zhang et. al., Nature \textbf{438}, 201 (2005);A. H. Castro
Neto \textit{et. al.,} \textit{\emph{Rev. Mod. Phys.}} \textbf{\textit{\emph{81}}}\textit{\emph{,
109 (2009).}}

\bibitem{stolyarova}E. Stolyarova \emph{et al.}, PNAS \textbf{104},
9209 (2007); G. M. Rutter \emph{et al}., Science \textbf{317}, 219
(2007); V. Brar \emph{et al}., Appl. Phys. Lett. \textbf{91}, 122102
(2007);M. Ishigami \emph{et al}., Nano Letters, \textbf{7}, 1643 (2007);Y.
Zhang \emph{et al}., Nat. Phys. \textbf{4}, 627 (2008);V. Geringer
\emph{et al.}, Phys. Rev. Lett. \textbf{102}, 076102 (2009);G. Li
\emph{et al}, Phys. Rev. Lett. \textbf{102}, 176804 (2009).

\bibitem{eigler}D. M. Eigler \textit{et al}, Nature \textbf{344},
524 (1990).

\bibitem{uchoa}B. Uchoa \emph{et. al.,} Phys. Rev. Lett. \textbf{101},
026805 (2008).

\bibitem{sengupta}K. Sengupta \emph{et. al.,} 
Phys. Rev. B \textbf{77}, 045417 (2008).

\bibitem{fradkin90} D. Withoff and E. Fradkin, Phys. Rev. Lett. \textbf{64},
1835 (1990); G.-M. Zhang \textit{et al.}, Phys. Rev. Lett. \textbf{86},
704 (2001); M. Hentschel \textit{et al.}, Phys. Rev. B \textbf{76},
115407 (2007), B. Dora \textit{et al.}, Phys. Rev. B \textbf{76},
115435 (2007).

\bibitem{fano}U. Fano, Phys. Rev. \textbf{124}, 1866 (1961).

\bibitem{kras} A. V. Krasheninnikov \textit{et al.}, Phys. Rev. Lett.
\textbf{102}, 126807 (2009).

\bibitem{note2} The hybridization in Eq. (\ref{eq:baf}) can be easily
generalized for $d$ and $f$-wave representations of the localized
orbital, and also for substitutional impurity defects.

\bibitem{anderson61} P. W. Anderson, Phys. Rev. \textbf{124}, 41
(1961).

\bibitem{note}The non-linear corrections of the spectrum give a finite
but small renormalization of the level, $\epsilon_{0}$, for $|\omega|\lesssim t$.

\bibitem{plihal}M. Plihal \emph{et al.} 
Phys. Rev. B \textbf{63}, 085404 (2001).

\bibitem{manoharan}H. Manoharan \emph{et al.} (unpublished).

\bibitem{sengupta2}K. Saha \emph{et al.}, arxiv:0906.2788 (2009). 
\end{thebibliography}
\end{document}